\documentclass[11pt,epsf,a4paper]{article}

\usepackage{epsf, rotating}
\newcommand{\beq}{\begin{equation}}
\newcommand{\eeq}{\end{equation}}
\newcommand{\beqn}{\begin{eqnarray}}
\newcommand{\eeqn}{\end{eqnarray}}
\newcommand{\ra}{\mbox{$\rightarrow$}}
\newcommand{\xmn}{\mbox{$\chi_{\mu \nu}$}}

\begin{document}

\begin{titlepage}
{
\begin{flushright}
23 February 2000
\end{flushright}
\vspace{3.0cm}
\begin{center}
{\Large {\bf
Search for spatial anisotropy in $\beta$-decays
} } \\
\vspace{1.5cm}
{\Large E}UGENIY {\Large N}OVIKOV AND {\Large P}AVEL {\Large P}AKHLOV \\
\vspace{0.8cm}
{\large {\it 
Institute of Theoretical and Experimental Physics \\
B.Cheremushkinskaya, 25, 117259, Moscow, Russia } } \\
\end{center}

\vspace{3.0cm}

\noindent
{\bf Abstract.} 
Theoretically a possibility of covariance violation in weak decays 
is not ruled out from the first principles, while there are some models 
predicting non-covariance revealed in weak interactions. The experimental 
evidence for isotropy violation in $\beta$-decays was recently reported.
We present a study of the dependence of electron flow rate and 
$\beta$-electron energy in the decay of $Sr^{90}$ with respect to the 
direction of electron emission. An upper limit of $1.4 \cdot 10^{-5}$ 
on directional dependence of $\beta$-electron energy was obtained.
}

\end{titlepage}

\section{Introduction}

Possible non-covariance of laws of nature was searched intensively during 
this century. Experimentally no evidence of covariance breaking in
general relativity and electrodynamics was observed in the famous 
E\"otv\"os and Hughes-Drever experiments \cite{etw,drh}. These precise 
experiments narrowed the field of possible non-covariant models, 
but still left a room for them.

Non-covariance, in particular Lorentz non-invariance, 
was proposed to be searched in weak decays by R\'edei in 1966 \cite{red}. 
He extrapolated to weak decays the argument of Blokhintsev \cite{blok}, 
who pointed out that possible existence of the universal length parameter 
violates the Lorentz invariance by introducing a preferred Lorentz frame, 
where this parameter is measured. In R\'edei's approach the lifetime of weakly 
decaying particles turns out to be anomalously dependent on their boost:
\beq
\tau (v) = \gamma \tau_0 (1+ \gamma^2 a_0^2), 
\eeq
where $\tau_0$ is the lifetime in the preferred frame and 
$a_0$ is the universal length parameter. Experimentally an indication 
of the anomalous dependence of the lifetime of charged pions and kaons 
on their boost was observed, while a strict limit for such a 
dependence was obtained for muon decays \cite{pkm}. Since the covariance 
violation mechanism was supposed to be the same for muons and hadrons 
the latter measurements were considered to rule out R\'edei's hypothesis
at a high accuracy level.
 
In the end of 70-th the question of non-covariance in weak decays 
was renewed by Nielsen and Picek \cite{niel}. They proposed not a vector 
but a tensor covariance breaking mechanism by
introducing a non-covariant metric into the 
Higgs kinematic term $h^{\mu \nu} D_{\mu}\phi ( D_{\nu} \phi )^{*},$ 
where $h^{\mu \nu}$ is different from the metric tensor $g^{\mu \nu}$. 
Such a term results in non-covariant observables in weak decays, while 
electrodynamics and gravitational sectors remain covariant. Indeed, the
new metric is revealed in a gauge boson mass term 
($h^{\mu \nu} A_{\mu}^a A_{\nu}^a \langle 
\phi \rangle^2$) and results in a tensor structure of the Fermi 
constant: $G_F^{\mu \nu}=\frac{\sqrt{2}g_2^2}{8 h^{\mu \nu} 
\langle \phi \rangle^2} 
\simeq \frac{\sqrt{2}g_2^2}{8m_W^2} (g^{\mu \nu} +\xi^{\mu \nu})$.
In \cite{niel} the metric $\xi^{\mu \nu}$ is assumed to be 
isotropic but not Lorentz invariant. In the simplest form it is 
parametrized by a single parameter $\alpha$ ($\xi^{\mu \nu} =\alpha~
{\rm diag}(1,1/3,1/3,1/3)$). Under these assumptions the dependence of the 
lifetime of weakly decaying particles on their boost is derived. 
It turns out that the lifetime of charged pions and kaons is affected by the 
additional boost term in a way similar to the R\'edei approach, while for 
muons the effect cancels out.

In the last decade the arguments in favor of Lorentz invariance violation 
came from the string theory, where non-local objects -- strings can lead 
to the spontaneous breaking of the covariance \cite{string}. 
The covariance-violating term is generated when tensor rather than scalar 
fields gain the vacuum expectation values. If this tensor field couples to 
the weak gauge bosons we come to the approach of Nielsen and Picek.

The metric $\xi^{\mu \nu}$ being anisotropic leads to visible anisotropy 
in weak decays, which can be searched for experimentally. Assuming 
the simplest one-parameter case ($\xmn =\alpha~{\rm diag}(1,0,0,1)$) 
the flow rate of daughter particles in weak decays is calculated to be 
direction dependent. Consider for example muon decay  at rest.
\beqn
\mu^+(p) \ra e^+(k) + \nu_e (q_1) + {\bar{\nu}}_{\mu} (q_2).
\eeqn
Following calculations from \cite{niel} non-covariant term in muon 
differential width is equal to: 
\beqn
\nonumber
\frac{d\Gamma}{ dk d\cos{\theta} d\phi } =
\frac{G_F^2 k \xmn}{24\pi^4m} (2q^2p^{\mu}k^{\nu}+
(p\cdot k)q^{\mu}q^{\nu} - (k\cdot q) p^{\mu} q^{\nu} - \\
-(p \cdot q) k^{\mu} q^{\nu})  = 
\frac{G_F^2 k \alpha}{24\pi^4m} (m^3(m-3k) -
k^2m^2\cos(2\theta)),
\eeqn
where $q$ is the sum of the two neutrino momenta $q=q_1+q_2=p-k$ and 
$\theta$ is the angle of the electron momentum with respect to $z$-axis 
(further referred as ``preferred axis'').
The integration over electron momentum and polar angle $\phi$ 
gives: 
\beq
\frac{d\Gamma}{d\cos{\theta}} =
(1 + 2\alpha \cos{(2\theta)}) \frac{d\Gamma_{SM}}{d\cos{\theta}}. \\
\eeq

The $\beta$-decays of neutron and nuclei are calculated in a similar 
way. The $\beta$-electron rate exhibits a similar directional dependence:
\beq
\frac{d\Gamma}{d\cos{\theta}} =
(1 + A\alpha \cos{(2\theta)}) \frac{d\Gamma_{SM}}{d\cos{\theta}}, \\
\eeq
where $A$ is $\mathcal{O}(1)$ and depends on nuclear form-factors.

An indication of the anisotropy of the light propagation through the Universe 
published in \cite{nod} gives another argument in favor of isotropy
violation, though on macro scales. In particle physics an evidence of the 
directional dependence of the $\beta$-decay rate was reported recently 
\cite{dub}. In this paper we present our study of such a dependence. 
The upper limit for the spatial anisotropy obtained here is much stricter 
than the effect reported in \cite{dub}. 

\section{Experimental Setup}

The dependence of $\beta$-decay of $Sr^{90}$ on the direction of the 
electron emission was studied. To detect 
$\beta$-electrons two different options were used: scintillators viewed 
by photomultipliers and pad silicon detectors. The detectors were placed 
in front of a radioactive source of an intensity of $15~$mCi corresponding 
to $10^{8}$ decays per second at a distance of $4~$cm (figure~1).
\begin{figure}[htb]
\begin{picture}(100,204)
\put(0.0,0.0){\includegraphics{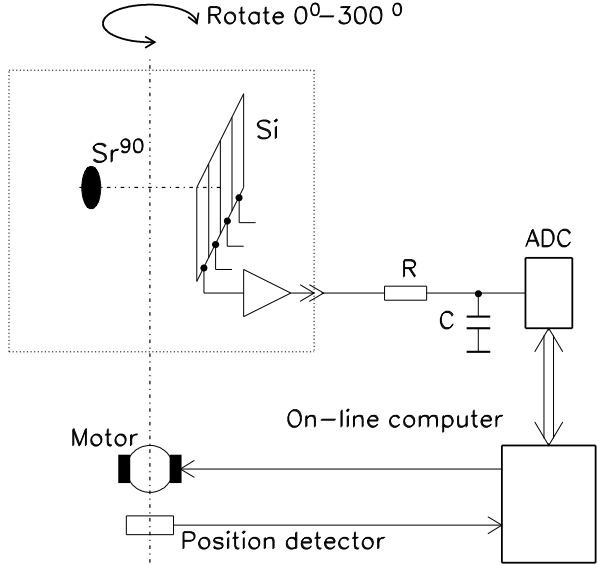}}
\end{picture}
\caption[aa]
{\small The sketch of the experimental setup.}
\label{pict:one}
\end{figure}

The plastic scintillator had an area of 12$\times20~$mm$^2$ 
and a thickness of $10~$mm. The {\it Hamamtsu-5600U} photomultiplier was 
attached to a side of the scintillator. Signal from the photomultiplier
exceeded the discriminator threshold of $30~$mV (corresponding to 
electron energy $\sim200~$keV) was counted by commercial CAMAC scaler
during some exposure time. The timing was provided by a quartz timing 
unit with a time stability of better than $10^{-5}$.

The silicon detector consisted of 4 rectangular pads with an area 
size of  $15\times20~$mm$^2$ and a thickness of $300~\mu$. Signals from 
the silicon pads were amplified by operational amplifiers. Under high 
electron rates with moderate time characteristics of the amplifier, 
we had to integrate the signals, rather than count them. Thus the 
$\beta$-electrons energy deposition was measured with silicon detector. 
Signals were integrated by RC-circuit with integration time of $250~\mu$sec 
and read out with a frequency of $5~$kHz by commercial ADC. The successive 
readings thus were not totally independent, their correlations
were taking into account while the data was analyzed.

In order to measure the decay rate in different directions, it was possible 
to use the daily rotation of the Earth, keeping the measuring device at a 
fixed position (as in \cite{dub}). However, in this case a very
high stability of the measurement conditions should be provided during 
days. High voltage and threshold potentials could change because of daily 
temperature and humidity variations resulting in a variation of the decay 
rate count with time which is a source of irreducible systematic error. 
The required stability 
seems unrealistic for the measurements with the accuracy of the order of 
$10^{-5}$. Therefore the decision was to rotate the experimental 
setup artificially with minimal period such that the environment parameters
could not change significantly. The Monte Carlo simulation
with the actual characteristics of our device shows that the rotation with a
period of a few minutes guaranteed the systematic changes of the count rate 
to be much smaller than the statistical error of its measurement.

The shielded source, detector, preamplifiers and power supply units were 
installed on the platform, rotated by an electric motor in the horizontal 
plane. The number of counts from the PM or the integrated charge 
from the silicon detector were measured during exposure time of 
$\sim 4~$sec, then the platform was rotated by $30^0$ and 
the measurement was repeated. The number of steps of single $30^0$ 
rotations in one direction was equal to 10 scanning the angle of $300^0$, 
then the measurements were repeated rotating in the opposite direction. 
The measurements were carried on continuously during 9 days. 

\section{Data analysis and results}
 
With the first option of the electron detector (photomultiplier) we faced
the irreducible source of systematic error. Since the amplification of the 
PM is influenced by the external magnetic field and the vector of the Earth 
magnetic field changed relative to the PM while rotated, we observed a
non-uniform behavior of the count rate at different directions. To reduce
this effect we used active compensation of the Earth magnetic field which
allowed to suppress it by a factor of $100$. The measurements showed 
that even with these special efforts, the observed nonuniformity of the PM 
counts was of the order of $10^{-4}$. The count rate dependence on the 
position of the rotating platform is shown in figure~2 for two hours of data 
\begin{figure}[h]
\begin{picture}(100,230)
\put(0.0,0.0){\includegraphics{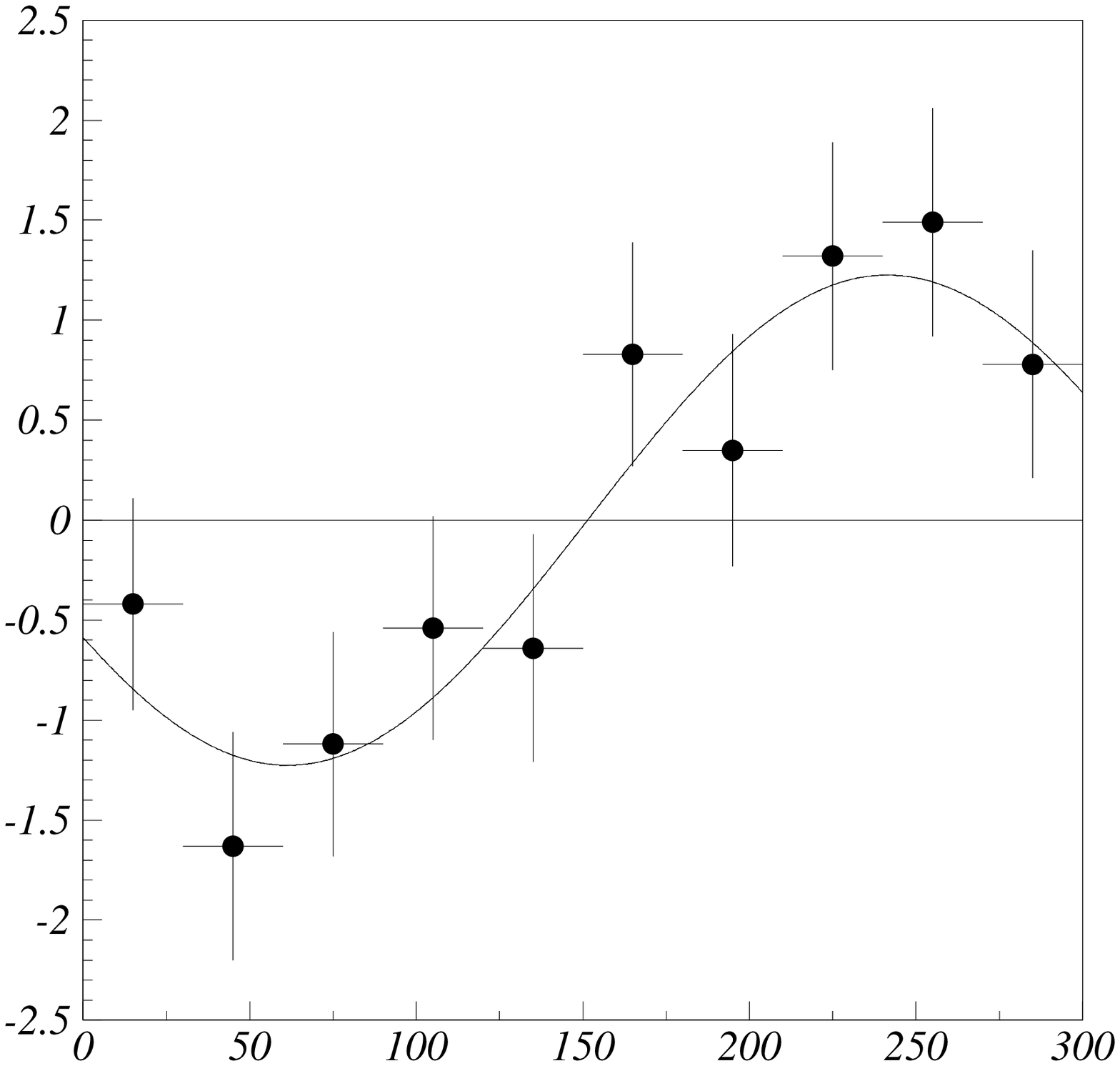}}
\put(47.0,220.0){\large $*10^{-4}$}
\put(150.0,-10.0){\large Angle of rotation~~~~~~~~~~ {\it [degree]}}
\end{picture}
\caption[aa]
{\small The relative difference of PM counting rate versus angle of rotation.}
\label{pict:two}
\end{figure}
taking. The well seen sinusoidal behavior is explained by the dependence of  
PM amplification on magnetic field projection. This was proved 
by changing the vector of the external magnetic field using the active 
magnetic compensation. Thus this option was used only for checking of our
sensitivity to the non-uniform effects.

The silicon detectors are not affected by the magnetic field at
the required level of accuracy. From the other hand the silicon detector and 
preamplifiers are much more sensitive to the temperature variations as 
demonstrated by figure~3. The integrated charge 
in the silicon detector depending on time during 9 days of data taking 
is presented in figure~3. The prominent periodical local maxima and minima
\begin{figure}[htb]
\begin{picture}(100,230)
\put(0.0,0.0){\includegraphics{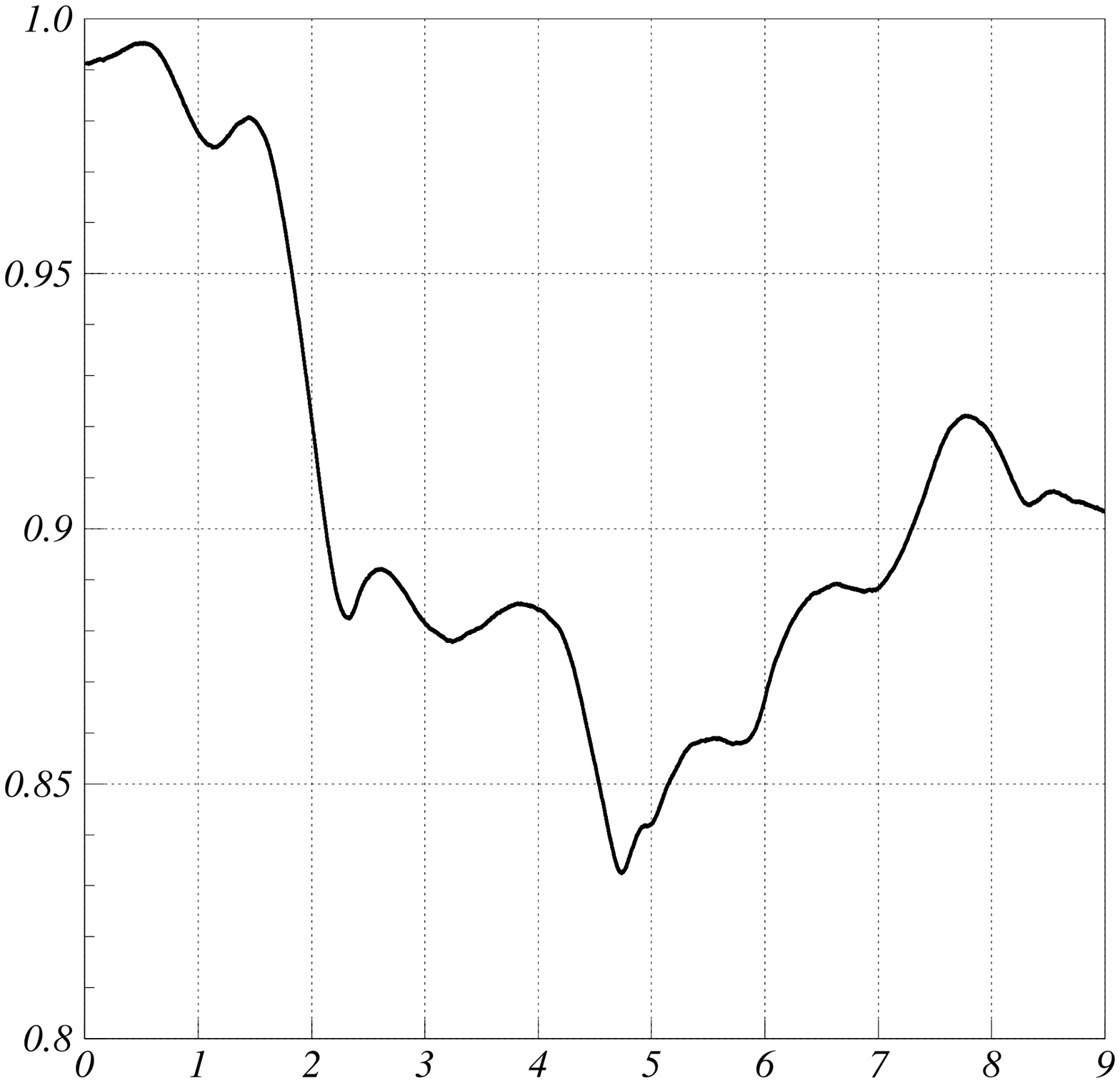}}
\put(15.0,210.0){\large Collected }
\put(22.0,196.0){\large charge }
\put(15.0,182.0){\large in silicon }
\put(0.0,168.0){{\it (abitrary units)}}
\put(192.0,-10.0){\large Time~~~~~~~~~~~~~~~~~~~{\it [days]}}
\end{picture}
\caption[aa]
{\small Time dependence of the collected charge in the silicon detector. 
}
\label{pict:three}
\end{figure}
are due to day-night temperature variation of about $3^0~$C. 
This instability resulted in only small error in 
non-uniformity measurements using the frequent rotations for different 
direction scanning.

The statistical error of the measurement of the integrated charge during one
step of measurement was calculated from the statistical fluctuations of two 
subsequent steps of measurements. The differences of values obtained in 
all pairs of successive measurements were plotted and than fitted with 
Gaussian function. For cross check we also derive the statistical error 
from the RMS of the ADC readings during one step of measurement, 
taking into account the correlation between two successive readings of 
charge from ADC (reading with interval of $200~\mu$sec while the 
integration time of RC-circuit is $250~\mu$sec). Both ways gave the similar 
values within $5\%$. Finally we confirm the extracted from the data values 
of the statistical errors by numerical calculations.
  
Each day of data taking was divided into 12 time intervals of $\sim 2~$hours. 
During each interval the Earth orientations was assumed to be the same. 
For each time interval the charge collected on four silicon pads for each 
orientation was summed up taking into account consecutive shifts in the 
position of each pad. Thus, the analyzed data contains information from
10 points of different orientation of the device relative to the Earth 
and 12 points of different Earth orientations. The data is presented in 
the form of 12 histograms in figure~4. Each of 12 histogram contains the 
dependence of the collected charge on the angle of the rotation 
of the device for some Earth position. All distributions was normalized to 
unity and unity was then subtracted, thus demonstrating only the net studied 
effect. Nowhere a signal of non-uniformity is seen.
\begin{figure}[htb]
\begin{picture}(170,350)
\put(0.0,0.0){\includegraphics{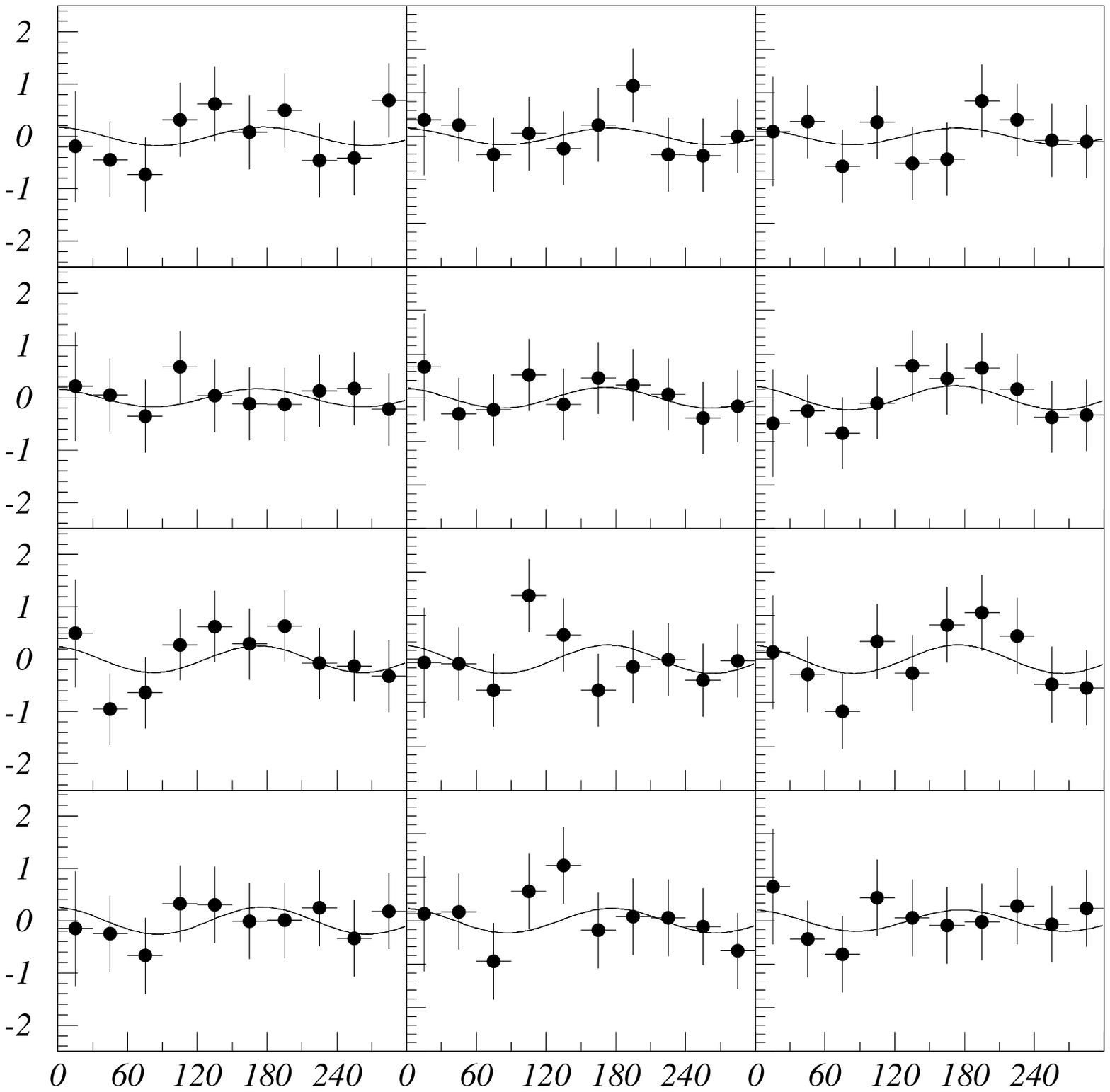}}
\put(10,340.0){\large $*10^{-5}$}
\put(140.0,335.0){\bf $I$}
\put(243.0,335.0){\bf $II$}
\put(346.0,335.0){\bf $III$}
\put(140.0,257.0){\bf $IV$}
\put(243.0,257.0){\bf $V$}
\put(346.0,257.0){\bf $VI$}
\put(140.0,179.0){\bf $VII$}
\put(243.0,179.0){\bf $VIII$}
\put(346.0,179.0){\bf $IX$}
\put(140.0,101.0){\bf $X$}
\put(243.0,101.0){\bf $XI$}
\put(346.0,101.0){\bf $XII$}
\put(160.0,0.0){\large Angle of rotation ~~~~~~~~~~~~~~~~~~~~~ $[degree]$}
\end{picture}
\caption[aa]
{\small Relative difference in the collected charge in the silicon detector 
versus the angle of rotation for the 12 ({\bf I-XII}) intervals of time. 
The solid line represents the best fit curve.}
\label{pict:four}
\end{figure}

The orientation dependent behavior of the $\beta$-electron energy deposition 
assuming the model discussed in the introduction
section is the following:
\beq
\frac{{\rm d}N}{{\rm d}\cos{\theta}}\sim 1+A(t)
\cdot \cos{(2\theta+\phi_0(t))}, 
\eeq
where the angle $\theta$ is the angle of the rotation of our device 
with respect to the axis south-north. The amplitude $A(t)$ and the 
phase of the cosine $\phi_0(t)$ change with time because of the Earth
rotation as shown in figure~5. 
\begin{figure}[h]
\begin{picture}(100,218)
\put(0.0,0.0){\includegraphics{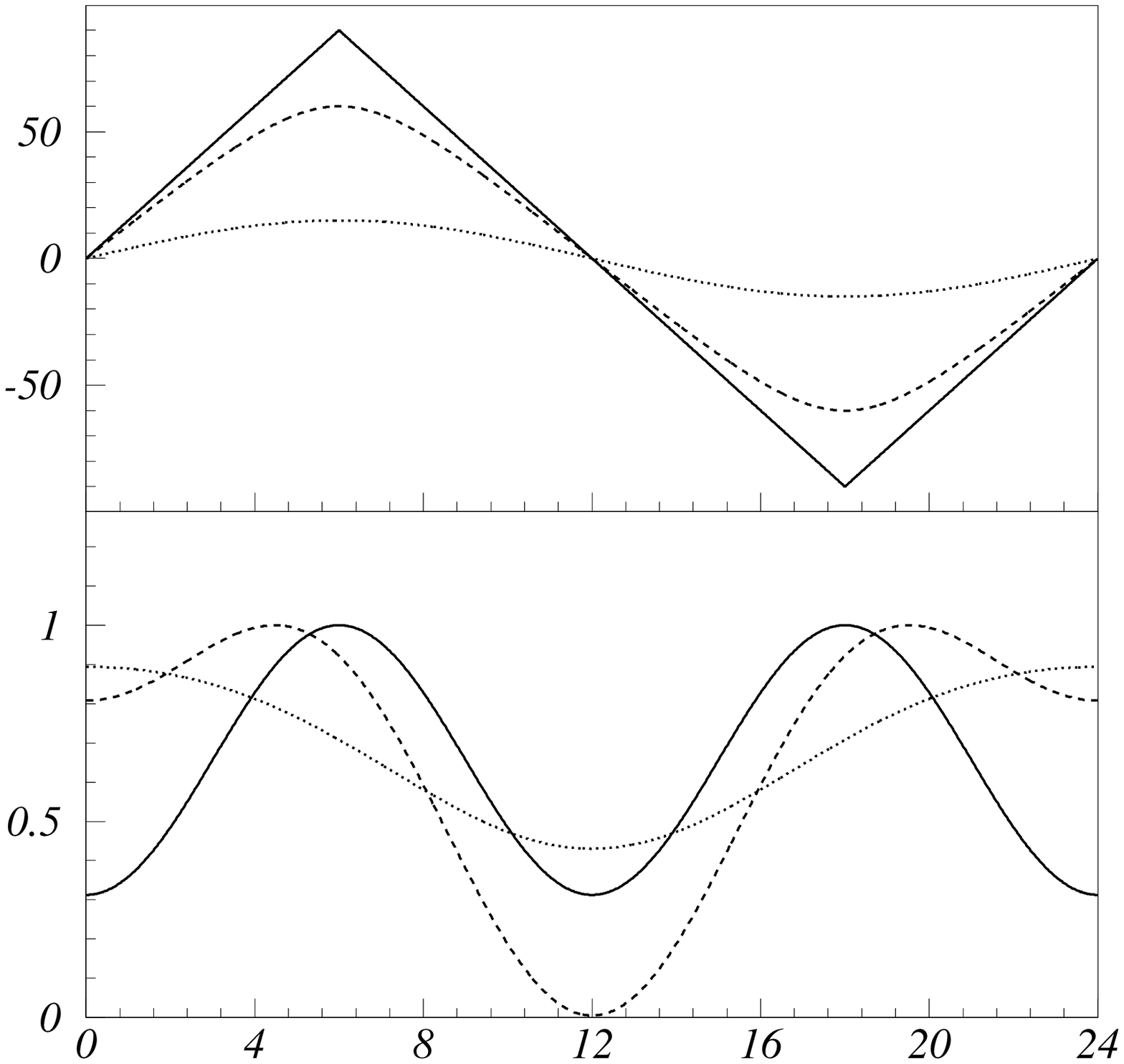}}
\put(288.0,195.0){\large a)}
\put(288.0,92.0){\large b)}
\put(47.0,200.0){\large $\phi_0(t)$}
\put(39.0,186.0){\large {\it [degree] }}
\put(50.0,100.0){\large $A(t)$}
\put(183.0,-10.0){\large Time ~~~~~~~~~~~~~~~~{\it[hours]}}
\end{picture}
\caption[aa]
{\small Time dependence of the initial phase $\phi_0(t)$ (a) 
and the amplitude $A(t)$ (b) from the formula (6) on time for the latitude 
of Moscow. The solid, dashed and dotted lines are plotted for the angle 
between $z$-axis and the axis of 
the Earth rotation equal to $90^0$, $60^0$ and  $15^0$, respectively.}
\label{pict:five}
\end{figure}
Different lines represent $A(t)$ and $\phi_0(t)$ behavior for different 
angles between the axis of the Earth rotation and the preferred axis for 
the latitude of the place of the experiment. One concludes that whatever 
the orientation of the preferred axis in the Universe, the non-uniformity 
of the electron flow is visible at least sometimes during the 
twenty four hours.

To set an upper limit for the dependence of electron energy on direction, 
these 12 histograms were fit simultaneously by a function with three free 
parameters: amplitude of the effect and two variables to describe direction 
of the preferred vector in the Universe. For each of 12 histograms the 
fitting function corresponds to the formula (6), while $A(t)$ and 
$\phi(t)$ are different for different histograms and are functions of three 
parameters described above.

The fit gives the value $A=(6.7\pm 3.6) \cdot 10^{-6}$ for the amplitude 
of the effect. The upper limit, derived from 
likelihood function, is calculated to be $1.4 \cdot 10^{-5}$ at the 
$90\%$ confidence level.

\section{Summary}

Dependence of the rate and energy deposition of 
$\beta$-electrons from $Sr^{90}$ decays on the direction of 
emission was investigated. Unlike the previously stated 
evidence for such a dependence in \cite{dub}, no signal of the 
non-uniformity in $\beta$-electron flow was observed. 
The upper limit on the amplitude of non-uniform behavior 
of $1.4 \cdot 10^{-5}$ was obtained. 

\vspace{1.0cm}
\noindent 
{\bf {\large Acknowledgments}}

It is our great pleasure to thank M.Danilov for support, useful 
discussions and interest to the work. We are very grateful to 
K.Boreskov, L.Okun and M.Voloshin for numerous theoretical discussions. 
We thank I.Tikhomirov, L.Laptin and A.Petryaev for their help in preparation
of the experiment.

\end{document}